\documentclass[aps,prl,twocolumn,groupedaddress,showpacs]{revtex4}
\usepackage[dvips,final]{graphicx}
\usepackage{color}

\def\TT{\mathcal{T}}
\def\hQ{\hat{Q}}
\def\hI{\hat{I}}

\def\ua{\uparrow}
\def\da{\downarrow}

\begin{document}

\title{Inelastic Backaction due to Quantum Point Contact Charge Fluctuations}

\author{C. E. Young}
\affiliation{Department of Physics, McGill University, Montreal, Quebec, Canada H3A 2T8}
\author{A. A. Clerk}
\affiliation{Department of Physics, McGill University, Montreal, Quebec, Canada H3A 2T8}

\date{\today}

\begin{abstract}
We study theoretically transitions of a double quantum-dot qubit caused by nonequilibrium charge fluctuations in a 
nearby quantum point contact (QPC) used as a detector.  We show that these transitions are related to the fundamental Heisenberg backaction associated with the measurement, and use the uncertainty principle to derive a lower-bound on the transition rates.  We also derive simple expressions for the transition rates for the usual model of a QPC as a mesoscopic conductor, with screening treated at the RPA level.  Finally, numerical results are presented which demonstrate that the charge noise and shot noise backaction mechanisms can be distinguished in QPCs having nonadiabatic potentials.  The enhanced sensitivity of the charge noise to the QPC potential is explained in terms of interference contributions similar to those which cause Friedel oscillations.
\end{abstract}


\maketitle

A fundamental consequence of the Heisenberg uncertainty principle is that the act of measurement necessarily perturbs the system being measured; these measurement-induced disturbances are termed ``backaction".  Research on quantum electronic systems has progressed to the point where backaction effects, often near the limits imposed by the uncertainty principle, are of key relevance to experiment \cite{Devoret:2000p734}. They are particularly important in experiments involving solid state qubits, as they set a limit on how well one can read out the qubits. In qubit-plus-detector systems, detector backaction can both dephase the qubit (i.e.~$T_2$ effects), as well as drive transitions between the qubit energy eigenstates (i.e.~$T_1$ effects, or ``inelastic" backaction).  The study of backaction is also of purely fundamental interest, as it can shed light on the physics of the detector, which is often an interesting nonequilibrium quantum system in its own right.

In this paper, we focus on the inelastic backaction associated with quantum point contact (QPC) charge detectors \cite{Field:1993p324}.  QPCs are among the most versatile and widely used charge detectors in quantum electronics; having a complete understanding of their backaction is therefore highly desirable. They are the standard readout for both charge and spin quantum-dot qubits \cite{Petta:2004p241,Elzerman:2004p1582}, and have also been used to make sensitive measurements of nanomechanical oscillators \cite{Poggio08}.  The backaction of QPC detectors has been the subject of considerable study. Theoretically, QPC-induced dephasing has been studied in Refs.~\cite{Levinson97,Aleiner97,Gurvitz:1997p51}, and measured experimentally in seminal experiments on mesoscopic interferometers \cite{Buks:1998p47}.  Theory has also shown that the backaction dephasing induced by a QPC can reach the fundamental quantum limit set by the Heisenberg uncertainty principle \cite{Pilgram:2002p8,Clerk03,Korotkov}. Inelastic QPC backaction has also received attention. Aguado and Kouwenhoven \cite{Aguado00} discuss such effects in a system where a double quantum-dot is measured by a QPC. They focus exclusively on a mechanism whereby QPC current flucutations (shot noise) lead to potential fluctuations across the double dot, which in turn drives transitions in the qubit.  Several recent experiments have attempted to investigate this inelastic backaction mechanism \cite{Onac:2006p757, Khrapai:2006p353, Gustavsson:2007p22, Gustavsson:2008p60}. 

Here we focus on an alternate mechanism for inelastic QPC backaction which involves its nonequilibrium charge fluctuations, as opposed to its current fluctuations. Recall that the physics behind these two kinds of noise is very different. Current fluctuations are related to the partitioning of electrons into transmitted and reflected streams by the QPC scattering potential; this results in a binomial distribution of the time-integrated current through the QPC \cite{Levitov96}.  In contrast, charge noise results from fluctuations in how long electrons spend in the scattering region (i.e.~dwell-time fluctuations) \cite{Pedersen:1998p9}. Importantly, the current noise mechanism proposed in Ref.~\cite{Aguado00} can in principle be completely suppressed by appropriate circuit design. In contrast, the inelastic backaction arising from QPC charge fluctuations is the fundamental Heisenberg backaction: for a fixed measurement strength, it cannot be made arbitrarily small.

Using the uncertainty principle, we derive a rigorous lower bound on the charge noise inelastic backaction that exclusively involves directly measurable quantities. We also derive simple expressions for transition rates valid for the usual model of a QPC as a coherent mesoscopic conductor.  Our results suggest that the charge noise mechanism may already play a significant role in experiments.  Further, we show that charge fluctuations are more sensitive to the nature of the QPC potential than current noise:  while the current noise is simply determined by the QPC transmission $\TT$, the charge noise also depends on the adiabaticity of the QPC potential. As a result, it should be experimentally possible to distinguish the current and charge noise backaction mechanisms by their dependence on $\TT$.  We provide an explanation for this effect in terms of interference contributions similar to those that give rise to Friedel oscillations.

Note that other aspects of QPC charge noise backaction on qubits have been studied before. B\"uttiker and collaborators have studied inelastic backaction due to equilibrium charge fluctuations, as well as dephasing due to nonequilibrium charge noise \cite{Pilgram:2002p8,Buttiker:2005p3410}. While the general form of the nonequilibrium charge noise has been derived \cite{Pedersen:1998p9}, its direct role in generating qubit transitions at weak coupling has not to our knowledge been previously elucidated.  Related work has examined Fermi-edge physics in a qubit-QPC system \cite{NazarovQPC2007}, as well as an indirect mechanism where QPC charge noise, in consort with spin-orbit coupling, yields transitions in spin qubits~\cite{Borhani:2006p2007}.

 \begin{figure}
 \includegraphics[width=3in]{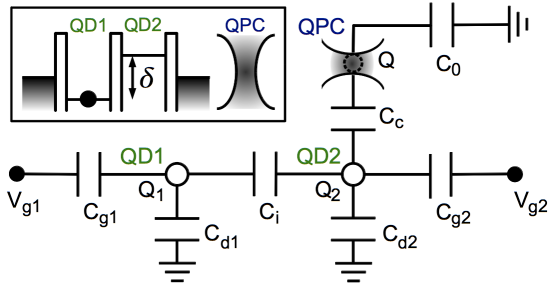}
 \caption{Electrostatic model of the coupled charge qubit-QPC system. Inset: Schematic showing the detuning $\delta$ of the DQD.}
 \label{system}  \end{figure}
\textit{Model.} While our results are general, we focus on the experimentally-relevant case of a double quantum-dot (DQD) charge qubit electrostatically coupled to a QPC charge detector (see Fig.~\ref{system}). The qubit Hamiltonian is $H_{DQD}=\frac{\delta}{2}\hat{\sigma}_z+t_c \hat{\sigma}_x$, where $\delta$ is the detuning and $t_c$ is the interdot tunneling amplitude. The eigenstates of $\hat{\sigma_z}$ correspond to an electron occupying either the left or right dot. We start by considering a completely general model of a spin-degenerate, single channel QPC with applied bias $V_{\rm qpc}$. The QPC conductance is $G=\frac{2e^2}{h}\TT$, where $\TT$ is the effective transmission.

The QPC acts as a detector by virtue of the fact that the potential felt by QPC electrons depends on the qubit's charge state. The qubit-QPC interaction Hamiltonian therefore necessarily includes a term of the form
\begin{equation}
	\hat{H}_{\rm int} = \frac{\overline{\Delta U}}{2} \hat{\sigma}_z \hat{Q}\, ; \hspace{0.5cm } 
	\hat{Q} \, \equiv \,\int d\vec{r} \; \frac{ \Delta U(\vec{r}) }{\overline{\Delta U} } \;\hat{\rho}(\vec{r}) .\label{Hint}
\end{equation}
Here, $\Delta U(\vec{r})$ is the change in the QPC scattering potential brought about by changing the qubit charge state, and $\hat\rho$ is the electron density operator of the QPC. For convenience, we introduce the parameter $\overline{\Delta U}$ representing the average value of $\Delta U(\vec{r})$ in the scattering region of the QPC. For the electrostatic model in Fig.~\ref{system}, and in the limit of weak coupling $C_c<<C_0$, we have $\overline{\Delta U}=\frac{2 C_c}{C_0} \frac{E_C}{e}$ where $E_C$ is the charging energy of the coupled right quantum-dot. Via $\hat{H}_{\rm int}$, the QPC potential and hence the conductance both depend on the qubit state; the QPC current $\hI$ therefore yields information regarding the qubit state. We assume throughout the realistic limit of weak coupling in which linear response theory applies. Thus the coupling $\hat{H}_{\rm int}$ changes the average QPC current by an amount $\delta \langle \hI(t) \rangle =\frac{\overline{\Delta U}}{2} \int dt' \lambda(t-t') \langle \hat{\sigma}_z(t') \rangle $, where $\lambda[\omega]= -\frac{i}{\hbar} \int_0^{\infty} \! d\tau e^{i\omega\tau} \langle \,[\hat{I}(\tau), \hat{Q}(0) ]\, \rangle$ is the finite-frequency gain of the detector. At $\omega=0$, the gain $\lambda[0]$ is simply related to $\Delta \TT$, the change in the transmission of the QPC brought about by changing the qubit state:
\begin{equation}
\delta \langle \hI \rangle = \langle \hat{\sigma_z} \rangle \frac{2e^2 V_{\rm qpc}}{h } \frac{\Delta \TT}{2} 
		\equiv \frac{\overline{\Delta U}}{2} \lambda[0] \langle \hat{\sigma_z} \rangle . \label{eq:Gain}
\end{equation}	
Eq.~(\ref{eq:Gain}) concisely expresses the fact that, via $\hat{H}_{\rm int}$, the QPC acts as a charge detector. This interaction is the fundamental qubit-detector coupling: without it one cannot use the QPC to measure $\hat{\sigma}_z$. Consequently, the backaction due to $\hat{H}_{\rm int}$ is the Heisenberg backaction associated with the measurement. Fluctuations in $\hQ$ both dephase the qubit and induce transitions between its eigenstates: it is the inelastic backaction that we wish to describe. As $\hQ$ is a weighted average of the QPC charge density, we refer to it as the QPC ``charge" in what follows.  

Before proceeding, note that one may also have a second qubit-QPC interaction of the form $H_{{\rm int},2} = e\hat{\sigma}_z \hat{U}_2(t)$, where $\hat{U}_2[\omega] \equiv Z_T[\omega] \hat{I}[\omega]$, and $Z_T$ is a transimpedance that converts QPC current fluctuations into qubit potential fluctuations. Ref.~\cite{Aguado00} provides a comprehensive description of the qubit transitions induced by this coupling; Ref. \cite{Chudnovskiy:2009p3933} furthers this analysis by considering the effects of a spatially-dependent $Z_T$. Note however that $H_{{\rm int},2}$ does not contribute to qubit detection: linear response tells us that $H_{{\rm int},2}$ cannot change $\langle \hat{I} \rangle$ at low frequency. The backaction mechanism of Ref.~\cite{Aguado00} is therefore {\it not} the Heisenberg backaction of the QPC. Moreover, the backaction due to $H_{{\rm int},2}$ can be made arbitrarily small by designing the QPC-DQD circuit to suppress $Z_T$. The backaction due to fluctuations in $\hat{I}$ can therefore be eliminated without affecting the sensitivity of the detector.

\textit{Heisenberg bound}.
Fluctuations in $\hQ$ can drive transitions between the qubit eigenstates; for weak coupling, the Golden Rule rates for excitation ($\Gamma_{Q\uparrow}$) and relaxation ($\Gamma_{Q\downarrow}$) are given by $\Gamma_{Q\uparrow/\downarrow}[\Omega] = \left(\overline{\Delta U}/\hbar\right)^2 \left( \frac{ t_c}{\hbar\Omega} \right)^2 S_{QQ}[\mp \Omega]$, where $S_{QQ}[\omega] = 2 \int dt e^{i \omega t} \langle \hat{Q}(t) \hat{Q}(0) \rangle$ is the quantum noise spectral density of $\hat{Q}$ and $\hbar\Omega = \sqrt{\delta^2 + 4 t_c^2}$ is the qubit level splitting. As this backaction is due to the coupling used to make the measurement, one can use the uncertainty principle to rigorously derive lower bounds on the transition rates {\it without} having to specify a QPC model or the coupling potential $\Delta U(\vec{r})$. Using the Heisenberg inequality which bounds the symmetric-in-frequency noise of a linear detector \cite{Braginsky,Clerk:2008p699}, as well the relation between the asymmetric-in-frequency noise and dissipative response coefficients, we find
\begin{equation}
	\Gamma_{Q \uparrow / \downarrow} \geq
		\left( \frac{  t_c }{ \hbar \Omega } \right)^2 \! \left[
	\frac{ \left | \overline{\Delta U} \lambda[\Omega] \right |^2}{4 \overline{S}_{II}[\Omega]} 
	\mp \frac{(e \overline{\Delta U})^2}{2\pi\hbar^2 \Omega} 
\textrm{Re}\!\left\{\frac{R_K}{Z_{\rm in}[\Omega]}\right\} \right] \label{un2}
\end{equation}
Here, $R_K\equiv\frac{h}{e^2}$ and $\bar{S}_{II}[\omega]=\frac{1}{2}(S_{II}[\omega]+S_{II}[-\omega])$ is the symmetrized QPC current noise. The input impedance of the QPC as seen by the qubit is  $Z_{\rm in}[\omega] \equiv  -\hbar/\left(\omega \int_0^\infty d\tau e^{i \omega \tau} \langle  [\hat{Q}(\tau) , \hat{Q}(0)]  \rangle \right)$. As $Z_{\rm in}$ is directly related to the QPC charge-charge susceptibility, it can be measured by driving Rabi oscillations in the DQD and measuring the induced oscillations in the QPC charge $\hat{Q}$ using a second QPC, similar to the experiment of Ref.~\cite{GoldhaberGordonQPC}. The magnitude and phase of the resulting oscillations in $\hat{Q}$ provide a measure of $Z_{\rm in}[\omega]$.

The bound in Eq.~(\ref{un2}) is completely general, and does not rely on any assumptions regarding the QPC. Most importantly, all of the parameters on the right-hand side are intrinsic properties of the QPC and directly measurable. Note also that $Re\{Z_{\rm in}^{-1}\}$ vanishes as $\omega^2$ when $\omega \rightarrow 0$, reflecting the capacitive coupling between qubit and QPC.

{\it Backaction transition rates.} We now specialize and treat the QPC as a coherent scatterer, described by a scattering matrix $s$ with both time-reversal and inversion symmetries. The effects of screening in the QPC are described using the method developed by B\"uttiker and collaborators \cite{Pilgram:2002p8, Pilgram:2003p359, Buttiker:2005p3410}, which combines the circuit model in Fig.~\ref{system} with the random-phase approximation (RPA). Within this model, the spatial variations in the QPC charge and internal potential are taken into account by using a capacitive network to describe them \cite{ButtikerQPCCharge,EPAPS}.

In the presence of screening, the total (screened) change in QPC potential $\Delta U_{\rm tot}(\vec{r})$, resulting from a change in the qubit state, differs from the bare (i.e.~unscreened) potential change $\Delta U(\vec{r})$ appearing in $H_{\rm int}$.  It is $\Delta U_{\rm tot}(\vec{r})$, not $\Delta U(\vec{r})$, which determines the change in the QPC transmission $\Delta \TT$. Screening also modifies $S_{QQ}$. Importantly, within the RPA, one finds that $\Delta \TT$ and $S_{QQ}$ are both determined by the same parametric change of the scattering matrix, given by $\Delta s =  \int d\vec{r} \Delta U_{\rm tot}(\vec{r}) \left[ \delta s/\delta U(\vec{r})\right]$, where $\delta/\delta U(\vec{r})$ is a functional derivative with respect to the local QPC potential \cite{ButtikerJMathPhys}. Using this approach \cite{EPAPS}, we find that, even in the presence of screening, one can relate the qubit transition rates to $\Delta \TT$ without needing to know the precise form of $\Delta U_{\rm tot}(\vec{r})$. In the low-temperature limit $k_B T \ll \{e V_{\rm qpc}, \hbar \Omega\}$, and for $\hbar \Omega \leq e V_{\rm qpc}$, we find 
\begin{eqnarray}
	\Gamma_{Q\ua}[\Omega] & = &  
		\left[ \frac{t_c}{\hbar\Omega} \right]^2\! \frac{(\Delta \TT)^2}{\TT (1-\TT)}  \frac{eV_{\rm qpc}-\hbar\Omega}{4\pi\hbar} \label{gamQup} \\
	\Gamma_{Q\da}[\Omega] & = &  \Gamma_{Q\ua}[\Omega]  + \frac{1}{\pi} \left[  \frac{t_c}{\hbar\Omega}  \right]^2 \! 
		\frac{ \left( e \overline{\Delta U} \right)^2}{\hbar^2 \Omega}  {\rm Re}\!\left\{\frac{R_K}{Z_{\rm in}[\Omega]}\right\}.  
\quad \label{gamQdn}
\end{eqnarray}
Here we have taken $s$ and $\Delta s$ to be approximately energy-independent within the energy ranges of interest.

Eqs.~(\ref{gamQup}) and (\ref{gamQdn}) are the central results of this paper. They allow one to make a simple prediction for the charge noise-induced inelastic backaction using only experimentally-accessible QPC quantities. The excitation rate $\Gamma_{Q \uparrow}$ has a particularly simple form that depends only on the QPC bias $V_{\rm qpc}$, transmission $\TT$ and sensitivity $\Delta \TT$. This is in stark contrast to the current noise backaction rates discussed in Ref.~\cite{Aguado00}, where one must also estimate the value of a transimpedance that is difficult to measure. Simple estimates using Eq.~(\ref{gamQup}) and typical experimental values (e.g.~Ref.~\cite{Gustavsson:2007p22}) suggest the expected charge noise backaction is at least comparable in magnitude to the shot noise mechanism in recent experiments. We stress that screening only enters through the measurable quantities $\Delta \TT$ and $Z_{\rm in}$ (see also Ref.~\cite{Pilgram:2002p8}).

\textit{Charge Noise Backaction Signature.} Next we outline a way to distinguish the charge and current noise backaction mechanisms in experiment. Here we note that the shot noise mechanism of Ref.~\cite{Aguado00} yields a backaction excitation rate $\Gamma_{I \ua} \sim S_{II} \sim \TT (1-\TT)$.  Thus an observed dependence of $\Gamma_{\ua}$ on the QPC transmisison of the form $\TT\,(1-\TT)$ (and hence a maximum at $\TT=1/2$) is often regarded as a signature of the shot noise mechanism \cite{Onac:2006p757,Gustavsson:2008p60}.  From Eq.~(\ref{gamQup}), we find that in general, $\Gamma_{Q \ua}$ {\it will not} have the same dependence on $\TT$: the only case where it does is when $\frac{\Delta \TT}{\TT \, (1-\TT)}$ is a constant that does not change as $\TT$ and $\Delta \TT$ are simultaneously varied by changing gate voltages. As noted in previous work \cite{Pilgram:2002p8, Clerk03}, this condition is satisfied {\it if and only} if the QPC potential is adiabatic \cite{Glazman} and so is the qubit-QPC coupling (i.e.~the variation of $\Delta U(\vec{r})$ is smooth compared to the Fermi wavelength $\lambda_F$).  If these conditions are violated, $\Gamma_{Q \ua }(\TT)$ will not have its maximum at $\TT = 1/2$. This provides a way to distinguish the charge and current noise backaction mechanisms.

To test this assertion, we have numerically studied the charge noise associated with a model 2D QPC saddle point potential using a recursive Green function approach extended to include the calculation of local properties \cite{Metalidis:2005p351}. We consider the potential
\begin{equation}
U(x,y) = W\,e^{-4 x^2/w^2}\,y^2 \label{V}
\end{equation}
where $W$ is the quantity modified by the coupling to the DQD. The gating effect of the qubit is therefore modeled as a shift $U(x,y) + \Delta U(x,y) = (1 + \overline{\Delta U}/(a^2 W)) U(x,y)$, where $a$ is the lattice spacing. Note that our goal is not to advocate a particular coupling model, but rather to study the dependence of the charge noise on the QPC adiabaticity. In Eq. (\ref{V}), the adiabaticity of both the potential and the coupling is set by the barrier width $w$.

\begin{figure}
\includegraphics[width=3in]{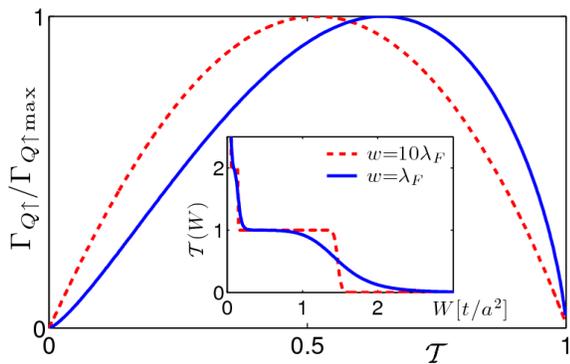}
\caption{Numerical results for the zero-frequency qubit excitation rate $\Gamma_{Q \ua}$ as a function of QPC transmission $\TT$. We use a lattice spacing $a\!=\!6.2{\rm nm}$, giving a hopping energy $t = \hbar^2/(2 m^* a^2)$ $=\! 14.5 {\rm meV}$, where $m^* \!= \! 0.068\, m_e$ in GaAs. For $E_F\!=\!16{\rm meV}$, the Fermi wavelength is $\lambda_F\!=\!6 a$. The dashed (red) line is for an adiabatic barrier with $w\!=\!10\lambda_F$, while the solid (blue) line is for a nonadiabatic barrier with $w\!=\!\lambda_F$; each curve is scaled by its maximum value. In the nonadiabatic case, $\Gamma_{Q \ua}$ is asymmetric and has its maximum at $\TT > \! 1/2$.  Inset: the corresponding transmission curves as a function of $W$.} \label{noise}
 \end{figure}

Fig.~\ref{noise} shows $\Gamma_{Q \ua}$ as a function of transmission. In the limit of an adiabatic potential and coupling  (i.e. $w\gg\lambda_F$), $\Gamma_{Q \ua}$ is to a good approximation proportional to $\TT \,(1-\TT)$, and therefore to $S_{II}$. In contrast, in the nonadiabatic case $w \sim \lambda_F$, the $\TT$-dependence of $\Gamma_{Q \ua}$ is markedly different: it is strongly skewed towards $\TT >1/2$. Observation of this asymmetry offers a way to distinguish the charge noise and current noise backaction.

\textit{Role of Friedel oscillations.} The increased charge noise backaction for $\TT > 1/2$ is due to interference contributions that are the nonequilibrium analogue of Friedel oscillations \cite{Friedel58}. In 1D, the continuity equation
\begin{equation}
\frac{d}{dt}\hQ = \frac{1}{\overline{\Delta U}} \int dx \left[ \frac{ d  }{d x} \Delta U(x) \right]  \hat{I}(x,t). \label{continuity}
\end{equation}
relates the charge operator $\hQ$ to a weighted average of the current flowing into the scattering region of the QPC. This is distinct from and independent of the transport current (flowing through the QPC) used to define the QPC conductance and shot noise $S_{II}$. The current operator $\hI$ of a coherent conductor contains spatially varying terms that oscillate with a wavelength $\lambda_F/2$.  These terms give rise to Friedel oscillations in the density-density correlation function, but do not contribute to, e.g., the average conductance or low-frequency shot noise \cite{Buttiker:1992p2642}.  In contrast, we find that they can enhance the nonequilibrium charge noise.  Their contributions to $S_{QQ}[\omega]$ via Eq.~(\ref{continuity}) correspond to interferences between incoming and transmitted waves, and thus scale as $\TT$.  If $\Delta U(\vec{r})$ varies slowly on the scale of $\lambda_F$, then the spatial averaging described by Eq.~(\ref{continuity}) strongly suppresses these terms. In the opposite limit, no such averaging occurs; the terms thus enhance the charge noise at higher values of $\TT$. Note that the same interferences yield the surprising result that the total charge in a perfect (i.e.~$\TT=1$), zero temperature 1D wire of length $L$ fluctuates; at $\omega=0$, the spectral density is $S_{QQ}[0] = (e^3 V \hbar / 4 \pi) \left[ \sin (k_F L )/ E_F \right]^2$, where $V\ll E_F/e$ is the bias and $E_F$ is the Fermi energy.

\textit{Conclusion.}  The fundamental Heisenberg backaction of QPC qubit detectors is due to nonequilibrium charge fluctuations. We have shown that the corresponding inelastic transition rates in the qubit can be expressed exclusively in terms of quantities that are directly measurable in experiment. We have also shown that, unlike the current noise backaction, the charge noise backaction need not be maximal at QPC transmission $\TT = 1/2$.

This work was supported by NSERC, FQRNT and CIFAR.

\bibliography{SQQbib}

\end{document}